\documentclass[aps,prl,twocolumn,showpacs,preprintnumbers,amsmath,amssymb,floatfix]{revtex4}

\usepackage{color}
\usepackage{graphicx}
\usepackage{bm}

\begin{document}
\title{Attractor Control Using Machine Learning}
\author{Thomas Duriez}
\email[]{thomas.duriez@univ-poitiers.fr}
\author{Vladimir Parezanovic}
\author{Bernd R. Noack}
\author{Laurent Cordier}
\affiliation{Institut PPRIME, CNRS - {Universit\a'e de Poitiers} - ENSMA, UPR 3346, D\a'epartement Fluides, Thermique, Combustion, CEAT, 43, rue de l'A\a'erodrome, F-86036 Poitiers Cedex, France}
\author{Marc Segond}
\affiliation{Ambrosys GmbH, Albert-Einstein-Str.\ 1-5, D-14469 Potsdam, Germany}
\author{Markus Abel}
\affiliation{Ambrosys GmbH, Albert-Einstein-Str.\ 1-5, D-14469 Potsdam, Germany}
\affiliation{LEMTA, 2 Avenue de la Forêt de Haye F-54518 Vandoeuvre-lès-Nancy Cedex, France}
\affiliation{University of Potsdam, Karl-Liebknecht-Str. 24/25 D-14476 Potsdam, Germany}


\date{\today}
\begin{abstract}
We propose a general strategy for feedback control design of complex dynamical systems exploiting the nonlinear mechanisms in a systematic unsupervised manner. These dynamical systems can have a state space of  arbitrary dimension with finite number of actuators (multiple inputs) and  sensors (multiple outputs). The control law maps outputs into inputs and is optimized with respect to a cost function, containing physics via the dynamical or statistical properties of the attractor to be controlled. Thus, we are capable of exploiting nonlinear mechanisms, e.g. chaos or frequency cross-talk, serving the control objective. This optimization is based on genetic programming, a branch of machine learning. This machine learning control is successfully applied to the stabilization of nonlinearly coupled oscillators and maximization of Lyapunov 
exponent of a forced Lorenz system. We foresee potential applications to most nonlinear multiple inputs/multiple outputs control problems, particulary in experiments. 
\end{abstract}
\maketitle

Non-equilibrium dynamical systems often show undesirable behaviour. Examples include fluid turbulence \cite{uriel1995turbulence,pope2000turbulent,eckhardt2007turbulence} which may adversely effect the forces on transport vehicles, financial crises with dramatic consequences for the world's economy \cite{guegan2009chaos}, or biophysical systems \cite{izhikevich2007dynamical} with obvious impact on our own life. Control can serve to stabilize nonlinear extended or discrete coupled systems, as lasers \cite{RevModPhys.77.783}, quantum systems \cite{PhysRevLett.111.103601}, or delayed feedback systems \cite{RevModPhys.77.783,doi:10.1142/S0217979212460071}. Consequently, the control of complex systems is an issue of major importance.

Cybernetics and control theory \cite{wiener1948cybernetics,astrom2010feedback,rowley2006dynamics,hoepffner2009input} have established a framework for control actions, mainly for stabilizing equilibria or reference trajectories; it is usually based on a linearization of the dynamical systems. In nonlinear dynamics, alternative strategies have been proposed, like time-delayed control with embedding, or synchronization \cite{Pikovsky-Rosenblum-Kurths-01,abel2009synchronization,gravier1999control,rosenblum1996phase} or stabilization of unstable periodic orbits \cite{ott1990controlling}. 

One major complication in extended systems lies in the role of dynamically destabilizing modes and their nonlinear interactions. Often, the cost function used to evaluate the action of the control quantifies a long-term property of the attractor, while the nonlinear control response becomes unpredictable after a much shorter prediction horizon. Thus, model-based control design becomes next to impossible. In particular, stabilization of equilibria is generally not doable in complex systems with limited control authority. 

Our approach hints to a pragmatic and fundamental solution out of this dilemma: we propose a  model-free control design using the tools from machine learning, in particular the genetic programming 
(GP)~\cite{koza1992genetic,koza1999genetic} as most suitable method. In general, machine learning comprises such important concepts as support vector machines~\cite{scholkopf2002learning}, neural networks~\cite{noriega1998direct}, or genetic algorithms to determine optimal parameters \cite{melanie_mitchell_book}. GP is a biologically inspired~\cite{wahde2008biologically} function optimization method. Here, GP is used to identify the optimum feedback law to control the properties of an attractor, focusing on strongly nonlinear dynamical systems. An art in GP lies in the  appropriate definition of  a cost function to be optimized. 

The applications lie dominantly in extended systems, as those mentioned above. Here, we describe the main physical and computational ideas using two examples: the first example is a generalized mean-field model with only 2 oscillating constituents, not controllable by linear methods. The second one is a forced Lorenz system by which we demonstrate the original use of the cost function: we want to maximize destabilization of the system within prescribed bounds. Its applications may lie in mixing systems, as in the case of combustion.

In the following, we restrict ourselves to ordinary differential equations, without loss of generality. The system is represented in phase space by the vector $\mathbf{a}\in \mathbb{R}^{n_a}$, it is measured by sensors $\mathbf{s}\in \mathbb{R}^{n_s}$, and controlled  by actuators $\mathbf{b}\in \mathbb{R}^{n_b}$,
\begin{equation}
\frac{d\mathbf{a}}{dt} = \mathbf{F} \left( \mathbf{a}, \mathbf{b}\right)\, ,\;
\mathbf{s}       = \mathbf{H} \left( \mathbf{a}\right)\, ,\;
\mathbf{b}  = \mathbf{K} \left(\mathbf{s}\right) \,,
\end{equation}
with $\mathbf{F}$ a general nonlinear function, $\mathbf{H}$ the measurement function, and $\mathbf{K}$ the sensor-based control law. This law shall minimize the state- and actuation-dependent cost function:
\begin{equation}
J = J(\mathbf{a},\mathbf{b}).
\end{equation}
The cost function value grades how a given control law $\mathbf{K}(\mathbf{s})$ performs relatively to the problem at stake. That function can be formulated in order to put the system in a desirable state as equilibrium (our first example) or in order to optimize a given measure on the system such as Lyapunov exponents (our second example). The lower the value of the cost function, the better the control law solves the problem, thus the cost function is a transcription of the control problem for the designated dynamical system.

\begin{figure}
\begin{tabular}{c}
\includegraphics[width=8cm]{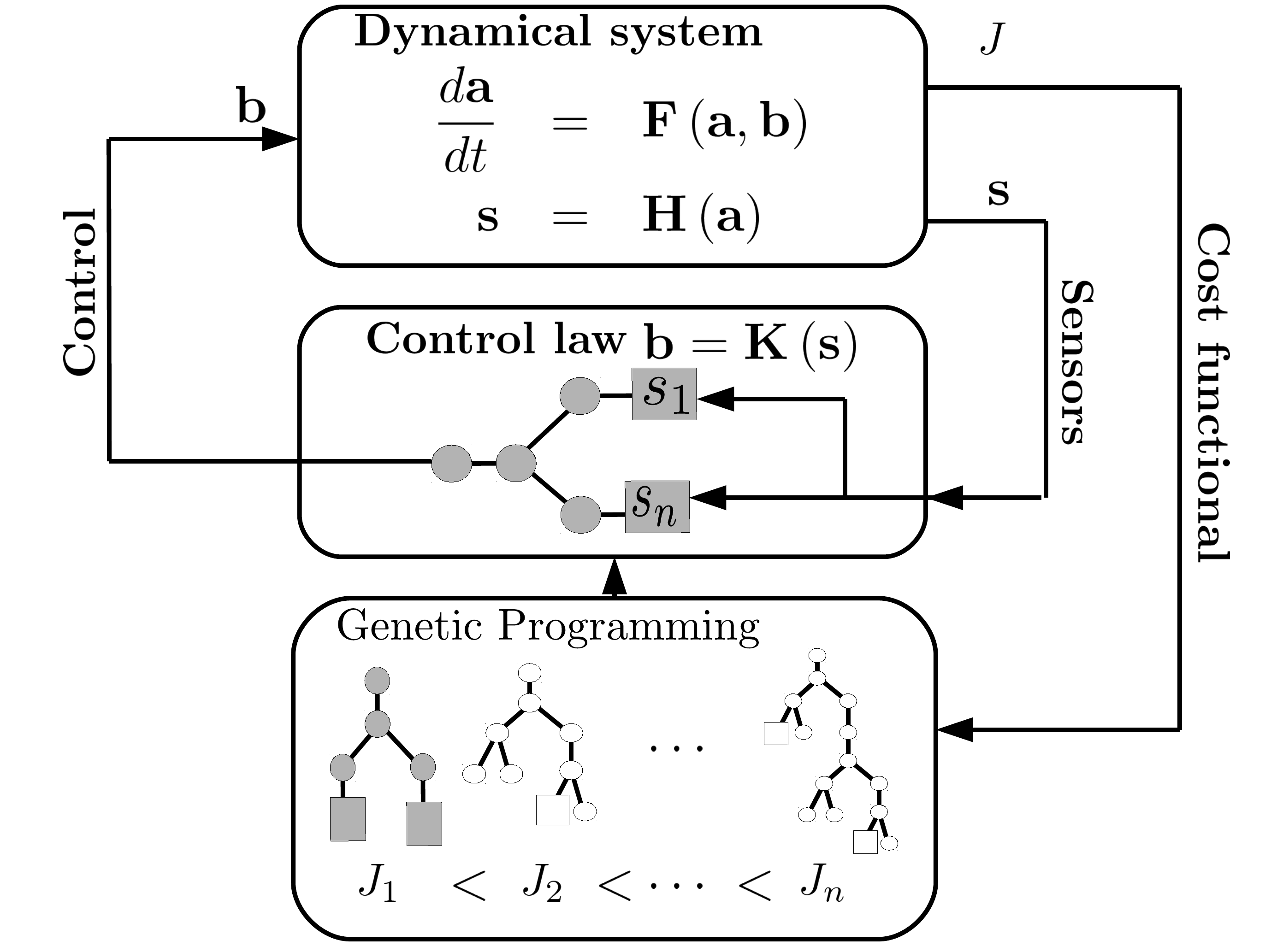}\\
\includegraphics[width=7cm]{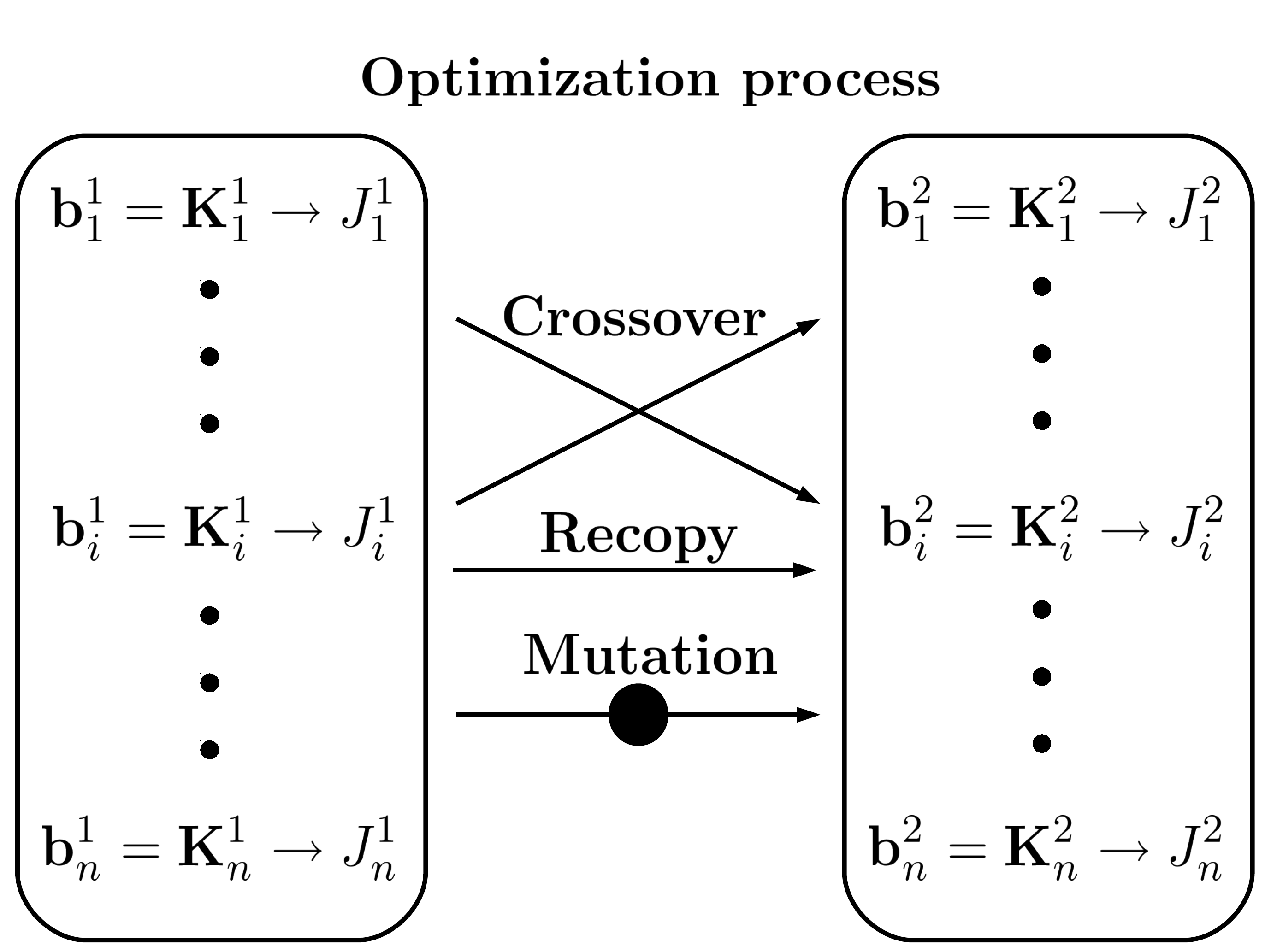}
\end{tabular}
\caption{Top: Control design using GP. During a learning phase, each control law candidate is evaluated by the dynamical system. This is iterated over many generations of individuals. At the end of the process, the best individual (in grey) is determined and used for control. Bottom: Production of a new generation of individuals: each individual $K_{i}^{m}$ is ranked by their cost, $J_{i}^{m}$, $i$ pointing to the $i^{th}$ individual, $m$ to the $m^{th}$ generation. An individual of the subsequent
generation can be a copy, a mutation or the result of the cross-over of individuals selected in the preceding generation, according to their cost.
} 
\label{fig:GPprocess}
\end{figure}

We propose a model-free design of the control law: our method integrates concepts of genetic programming into control of dynamical systems. The genetic programming is used to design the best control law $\mathbf{K(s)}$ as a composition of elementary functions. A first set of control law candidates (called individuals) is generated through random composition of the elementary functions. 
The exploited GP algorithm~\cite{ecj} combines these operations as a tree \cite{koza1999genetic}, which allows to generate any linear or nonlinear function as initial generation of individuals. Each individual is attributed a cost through the evaluation of $J(\mathbf{a},\mathbf{b})$. The next set  of individuals (called generation) is generated through mutation, cross-over or replication of individuals with specific rate for each process (Fig.~\ref{fig:GPprocess}). 

The individuals used to produce the new generation are selected based on how they minimize the cost function. A global extremum of the cost function is typically approximated well in a finite number of generation, if the population contains enough diversity to explore the search space. Though there is no general mathematical proof for convergence, the method has been proved to be successful~\cite{lewis1992genetic,nordin1997line}.

In our first study,  we consider a generalized mean-field model describing frequency cross-talk for a variety of physical phenomena including fluid flows~\cite{zielinska1997strongly,Luchtenburg2009jfm}. That model can be viewed as a generalisation of the Landau model~\cite{landau1975classical} for the phase transition from equilibrium to periodic oscillation. Since we focus on frequency cross-talk, we choose a simple form of this model with two oscillators, coupled through the parametric, nonlinear variation of one growth rate:
\begin{eqnarray}
\left[
\begin{matrix}
\frac{da_1}{dt}\\
\frac{da_2}{dt}\\
\frac{da_3}{dt}\\
\frac{da_4}{dt}
\end{matrix}\right]
= \left[\begin{matrix}
\sigma_1  & \omega_1 & 0 & 0 \\
-\omega_1 & \sigma_1 & 0 & 0 \\
0 & 0 & \sigma_2  & \omega_2 \\
0 & 0 & -\omega_2 & \sigma_2
\end{matrix}\right]
\left[
\begin{matrix}
{a_1}\\
{a_2}\\
{a_3}\\
{a_4}
\end{matrix}\right]
+
\left[
\begin{matrix}
0\\
0\\
0\\
b
\end{matrix}\right]\label{eq:GMM1}\\
\text{with }\sigma_1 = \sigma_{10} - (a_1^2 +a_2^2 + a_3^2 +a_4^2).\nonumber
\end{eqnarray}
Hereafter, we denote the sum of squared amplitudes as energy to avoid linguistic sophistication. We set $\omega_1=\omega_2/10=1$ and $\sigma_{10}=-\sigma_2=0.1$, so that the first oscillator, $(a_1,a_2)$, is unstable (would it be decoupled), while the other $(a_3,a_4)$ is stable. When uncontrolled ($b\equiv 0$), the nonlinearity drives the first oscillator to nonlinear saturation through the change of total energy. The actuation effects directly only the stable oscillator. This  system is arguably the simplest nonlinear dynamical system to exhibit frequency cross-talk. We choose to stabilize the first oscillator around its fixed point $(0,0)$ and thus a cost function which measures the fluctuation energy of that unstable oscillator. For any useful application, the energy used for control must be small, such that we penalize the actuation energy: 
\begin{equation}
J=\frac{1}{t_2-t_1}\int_{t_1}^{t_2-t_1}\left[ a_1^2(t) + a_2^2(t) + \gamma b^2(t)\right] \mathrm{d}t,
\label{eq:fitness}
\end{equation}
with $\gamma = 0.01$ as penalization coefficient. The quadratic form of state and actuation in the cost function is standard in control theory. Here, $t_{1}={5\tau}$ with $\tau=2\pi/\omega_1$ is chosen large enough to allow transients to decay and $t_2 = 100\tau$ is chosen in order to allow for meaningful statistics.

Knowing the nonlinearity at stake, an open-loop strategy can be designed: exciting the stable oscillator at frequency $\omega_2$ will provoke an energy growth which stabilizes the first oscillator as soon as $a_1^2 +a_2^2+a_3^2 +a_4^2 > \sigma_{10}$. It should be noted that linear control fails in using the frequency cross-talk mechanism. Indeed, the linearization of (\ref{eq:GMM1}) yields to two uncoupled oscillators. Thus, the first oscillator is uncontrollable using linear methods. 

We apply GP as a generic procedure with full-state observation ($\mathbf{s}\equiv\mathbf{a}$) in order to exploit all potential nonlinear mechanisms to control the unstable oscillator. In order to explore the function space, we use a set of elementary ($+,-,\times,/$) and transcendental (e.g. $\exp,\sin,\log_2$) functions. The functions are protected to allow them to take arguments in $\mathbb{R}$. Additionally, the actuation command is limited to the range $[-1\,,\, 1]$ in order to avoid too many integration failures and emulate an experimental actuator. Up to $50$ generations comprising $1000$ individuals  are processed. The algorithm is stopped only when the last generation is reached or if an evaluation yields $J=0$.

\begin{figure}[htp]
\includegraphics[width=8.6cm]{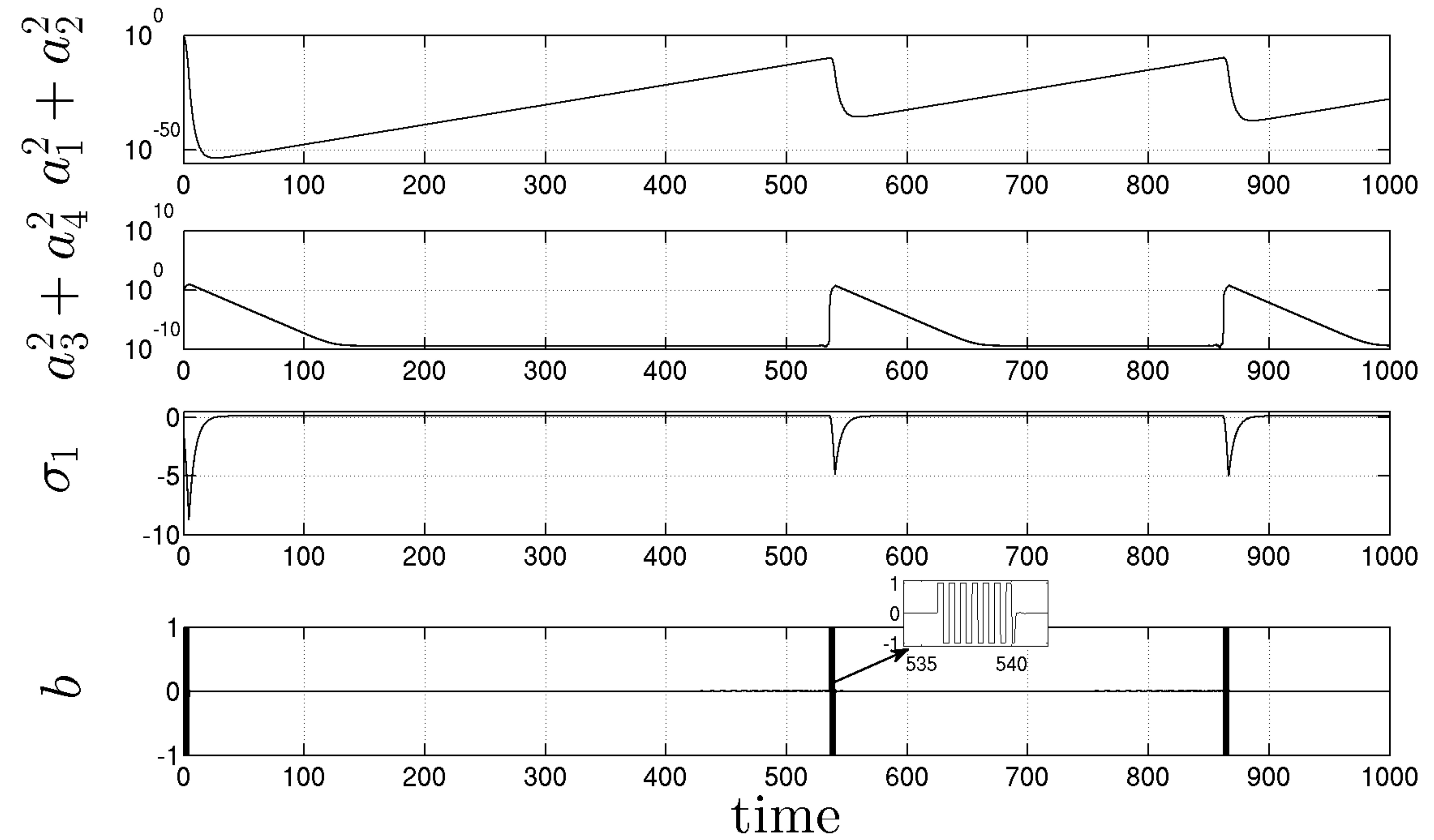}
\caption{Controlled generalized mean-field model. When the energy contained in the first oscillator (top) is larger than $10^{-10}$ the control (bottom) is exciting the second oscillator at frequency $\omega_2$, its energy grows so that $\sigma_1$ reaches $-5$. This results in a fast decay of the energy in the first oscillator after which the control goes in ``standby mode``. An animation of the controlled system can be found in~\cite{supplement}.}
\label{fig:energies}
\end{figure}

The control function ultimately returned by the GP process is a large expression, and due to the fact its argument is four-dimensional, it is hard to visualize. The full formula is given in \cite{supplement}. It can be summarized as follows:
\begin{equation}
b=K_1(a_4)\times K_2(a_1,a_2,a_3,a_4).
\end{equation}
The function $K_1(a_4)$ describes a phase control that destabilizes the stable oscillator. The function $K_2(a_1,a_2,a_3,a_4)$ acts as a gain based on the energies of the oscillators. The performance and behaviour of the control law is visible in Fig.~\ref{fig:energies}. The control law is energizing the second oscillator up to $10^0 \gg \sigma_{10}$, as soon as the first oscillator has an energy which is larger than $10^{-10}$. This is stabilizing the unstable oscillator very quickly, on a time scale of $10^{-5}$. When stabilization has happened, the control stays at very low values that keep the stable oscillator at a correspondingly low energy $\approx 10^{-10}$ while the energy of the unstable oscillator is exponentially growing close to its natural growth rate $\sigma_{10}$. The control exploits the frequency cross-talk and vanishes when not needed, i.e. $a_1\approx a_2\approx 0$. That control could not be found within a linear framework. In comparison with the best periodic excitation of the stable oscillator (with respect to $J$), less energy is used.

\begin{figure*}[htp]
\begin{center}
\includegraphics[width=0.9\textwidth]{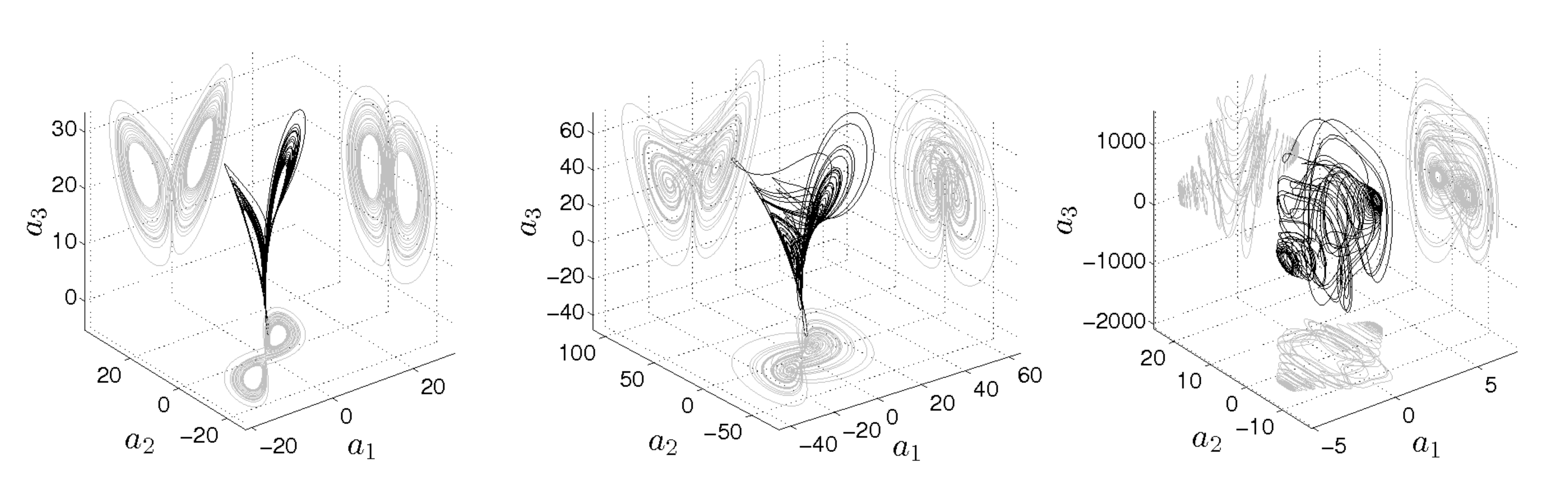}
\end{center}
\caption{Controlled Lorenz systems with $\sigma=10$, $\beta=8/3$ and $\rho=20$. Left: $\gamma=1$. The system exhibits chaotic behaviour ($\lambda_{1}=0.715$) close to the canonical chaotic Lorenz attractor with $\rho=28$ ($\lambda_{1}=0.905$). Center: $\gamma=0.01$. The system exhibits more complex trajectories, the nature of the central fixed point has changed and $\lambda_{1}=2.072$. Right: $\gamma=0$. The nature of all fixed points has changed. The non-penalization of the actuation leads to a change in the scales ($\lambda_{1}=17.613$). An animation of the controlled system can be found in~\cite{supplement}.}
\label{fig:3DLorenz}
\end{figure*}

As second example, we consider the Lorenz system \cite{Lorenz-63}, forced in the third component:
\begin{eqnarray}
\frac{da_1}{dt} &=&  \sigma\left(a_2 - a_1\right),\nonumber\\
\frac{da_2}{dt} &=& a_1\left(\rho-a_3\right)-a_2,\\
\frac{da_3}{dt} &=&  a_1a_2 - \beta a_3 +b,\nonumber
\label{eq:Lorenz} 
\end{eqnarray}
by the control law $b=K(a_1, a_2, a_3)$, which basically influences the growth rate of $a_3$. The Lorenz system can be stable, periodic or chaotic depending on the parameter set used. We employ $\sigma=10$, $\beta=8/3$ and $\rho=20$, such that the uncontrolled system ($b\equiv0$) is periodic. Instead of stabilizing an equilibrium, we demonstrate how to render the system chaotic, now. Existing strategies may stabilize or destabilize periodic orbits~\cite{ott1990controlling,pyragas1992continuous,schusterhandbook,janson2004delayed}. To reach chaotic behaviour, we aim at maximizing the largest Lyapunov exponent $\lambda_{1}$, again penalizing the actuation as a quadratic term $b^2$ with a factor $\gamma$. If $\lambda_{1}$ is positive, the system is chaotic and well-mixing. When considering a physical model the system needs to be bounded, i.e. the sum of the Lyapunov exponents must be negative. As a suitable cost function to be minimized and render the system chaotic we define:

\begin{equation}
\begin{array}{l l l l}
J   & =           & \exp(-\lambda_{1}) +{\displaystyle\frac{\gamma}{T}\int_{0}^{T}} b^2(t)\, \mathrm{d}t  &\quad\text{if}\,\, \sum_{i=1}^3 \lambda_i < 0,\\
J  &\rightarrow & \infty&\quad\text{if}\,\, \sum_{i=1}^3 \lambda_i \geq 0,
\end{array}
\end{equation}
where $T=100$ is the integration time and $\lambda_1\ge\lambda_2\ge\lambda_3$ are the Lyapunov exponents. 
These exponents are obtained by a standard algorithm~\cite{wolf1985determining,Kantz-Schreiber-97}. When the system is not bounded (i.e. it exceeds the bound we set in our program), $J$ is assigned the largest real number possible on the computer. The control law is based on the full state and the basic operations that compose $K$ are $+$, $-$, $\times$, $/$, as well as randomly generated constants. The maximum number of generations is again 50 with 1000 individuals each. To illustrate how the cost function definition influences the problem solved, we consider for $\gamma$ the values of $\gamma_1=1$, $\gamma_2=0.01$ and $\gamma_3=0$. After 50 generations, the best individuals \cite{supplement} have maximum Lyapunov exponents of respectively $\lambda_{1}=0.715$, $2.072$ and $17.613$. The changes in the system and the control function are displayed in Fig.~\ref{fig:3DLorenz}. The control laws associated with $\gamma_1$ and $\gamma_2$ cases are affine expressions of $a_3$, the diminution of the actuation cost leads to larger amplitude of the feedback. In those cases the most efficient controls lead the system into behaviours close to the canonical Lorenz system ($\rho=28$, $\lambda_{1}=0.905$). When $\gamma=0.01$ the nature (from saddle point to spiral saddle point) and position of the central fixed point are changed. When the actuation is not penalized ($\gamma=0$) the feedback law is a complex and fully non-linear law of all states. The nature and position of all the fixed points are changed as $\lambda_{1}$ reaches higher values. To our knowledge, no other model-based or model-free approach has been proposed in the litterature to optimize the largest Lyapunov exponent and controlling the attractor to reach a chaotic state.

Both examples illustrate how GP is progressing toward the minimum of the cost function. The statistical process that selects the individual for breeding allows individuals which are not optimal to be selected. This keeps diversity in the population and ensures that the GP process is not confined in a local minimum. For the stabilization of the mean-field model, the GP stopped after 35 generations, both oscillator energy and control energy vanished below  numerical accuracy of the integration scheme. If the sensors define a subspace of $\mathbb{R}^{n_a}$, reaching this result is not guaranteed, as the controller needs to modulate the destabilizing feedback on the oscillator ($a_3$,$a_4$) by monitoring the energy in both oscillators. 


\begin{figure}[htp]
\begin{center}
\includegraphics[width=8.6cm]{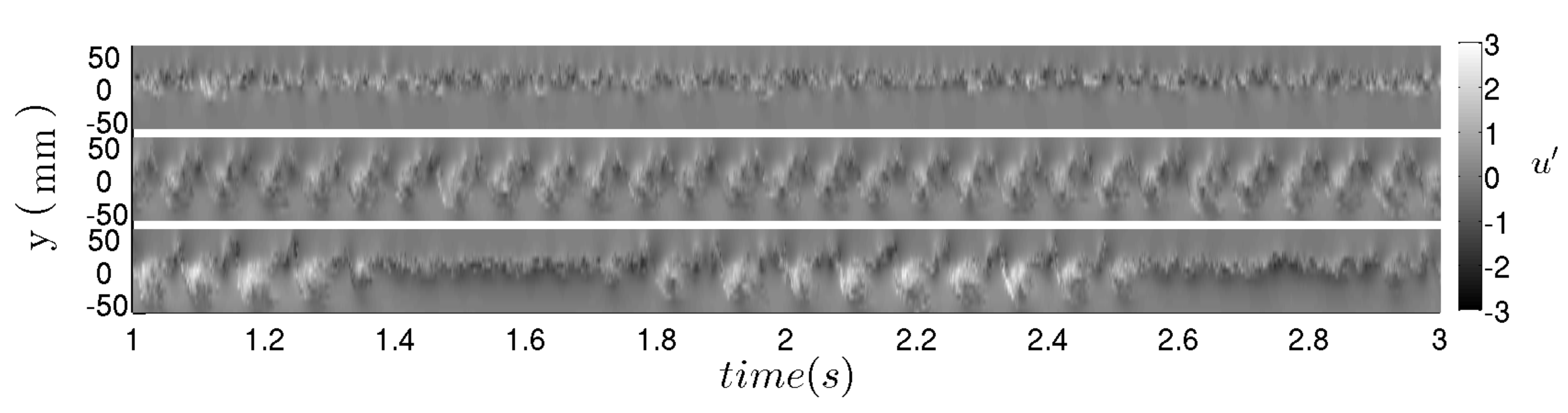}
\end{center}
\caption{Unforced (top), periodically forced (middle), and GP based closed-loop controlled (bottom) flows in the TUCOROM experimental mixing layer demonstrator~\cite{Parezanovic2013tsfp}. 24 hot-wires probes record the velocity fluctuations accross the shear flow forced by micro-pulsed jets at the tip of the separating plate. Machine learning control allows to reach fluctuation levels 3 times as high as the natural flow and 1.5 time as high as the best tested open- and closed-loop strategies. It introduces unexpected frequencies like the $\approx 1Hz$ modulation visible on the bottom figure. Details on the running experiment in~\cite{supplement}.}
\label{fig:chex}
\end{figure}

We have demonstrated a way to determine the optimum control of a complex dynamical system in a model-free framework. 
The stabilization of the mean-field model shows that we can obtain closed-loop control exploiting frequency cross-talk. This is of primordial importance for large-scale turbulence control featuring this frequency cross-talk. Currently, an experiment on an actuated turbulent shear layer is run with the proposed machine learning control strategy. At this moment, the achieved mixing enhancement is of 1.5 times larger than any tested open- or closed-loop method (Fig.~\ref{fig:chex}). Similarly, we have implemented a companion 2D DNS simulation with very promising results. This novel approach exhibits a high flexibility both in the class of systems it can address and in the specific problem it can solve exploiting the model-free formulation. Though a model is not needed, the more we know about the system, the better we can design the cost function according to the underlying physics. The major drawback of the model-free approach lies in the evaluation time, as each individual needs a simulation or experiment to be run. This translates in large time requirement should the process be serial. Consequently, in real applications, massive parallelization of computations or experiments will probably be needed. The relation of tree depth, number of generations, number of individuals with convergence is subject of ongoing research and may boost the performance considerably. The model-free control design is particulary interesting for experimental applications for which a model might not even be known - think about climate control or control of financial systems. 

We acknowledge funding of the French ANR (Chaire d'Excellence TUCOROM), also, MS and MA acknowledge the support of the LINC project (no. 289447) funded by EC’s Marie-Curie ITN program (FP7-PEOPLE-2011-ITN).

\end{document}